\documentclass[conference]{IEEEtran}
\IEEEoverridecommandlockouts
\usepackage{amsmath,amssymb,amsfonts}
\usepackage{algorithmic}
\usepackage{graphicx}
\usepackage{textcomp}
\usepackage{xcolor}
\def\BibTeX{{\rm B\kern-.05em{\sc i\kern-.025em b}\kern-.08em
    T\kern-.1667em\lower.7ex\hbox{E}\kern-.125emX}}
\usepackage[acronym]{glossaries}    

\usepackage[style=ieee]{biblatex} 
\makeglossaries
\newacronym{nft}{NFT}{Non-fungible Token}
\newacronym{ml}{ML}{Machine Learning}
\newacronym{dl}{DL}{Deep learning}
\newacronym{ai}{AI}{Artificial Intelligence}
\newacronym{nlp}{NLP}{Natural Language Processing}
\newacronym{erc}{ERC}{Ethereum Request for Comments}
\newacronym{lstm}{LSTM}{Long short-term memory}
\newacronym{api}{API}{Application Programming Interface}
\newacronym{recsys}{RecSys}{Recommendation System}
    
\begin{document}

\title{An Analysis of the Features Considerable for NFT Recommendations\\
}

\author{\IEEEauthorblockN{Dinuka Piyadigama}
\IEEEauthorblockA{\textit{Computer Science and Engineering} \\
\textit{University of Westminster}\\
London, UK \\
drpiyadigama@gmail.com}
\and
\IEEEauthorblockN{Guhanathan Poravi}
\IEEEauthorblockA{\textit{Department of Computing} \\
\textit{Informatics Institute of Technology}\\
Colombo 06, Sri Lanka \\
guhanathan.p@iit.ac.lk}
}

\maketitle

\begin{abstract}
This research explores the methods that \gls{nft}s can be recommended to people who interact with \gls{nft}-marketplaces to explore \gls{nft}s of preference and similarity to what they have been searching for. While exploring past methods that can be adopted for recommendations, the use of \gls{nft} traits for recommendations has been explored. The outcome of the research highlights the necessity of using multiple Recommender Systems to present the user with the best possible \gls{nft}s when interacting with decentralized systems.
\end{abstract}


\section{Introduction}
In recent months, the \Gls{nft} market has been growing exponentially as it appears to be the most widely accepted business application of Blockchain technology \autocite{dowling_is_2021}, since the introduction of crypto. With more and more people expected to enter connected digital environments such as the meta-verse \autocite{casey_newton_mark_2021}, it is clear that \gls{nft}s will play a huge role in tomorrow's internet \autocite{peter_allen_clark_what_2021} due to its ability to make digital items have scarcity, uniqueness, and proof of ownership, similar to physical items \autocite{noauthor_non-fungible_nodate}. Human interactions of the next decade on the internet may entirely rely on \gls{nft}s.

\subsection{What are \gls{nft}s?}
\Gls{nft}s are provably scarce unique digital assets that can be used to represent ownership \autocite{noauthor_erc-721_nodate}.
They can be one-of-a-kind rare artworks, collectible trading cards, and other assets with the potential to increase in value due to scarcity \autocite{conti_what_2021, fairfield_tokenized_2021}. While being digital assets, they also can be used to represent physical assets. A digital certificate of land/ qualification can be identified as a couple of examples. The biggest winners in the NFT space over the last few months have been digital artists who were able to sell art worth over \$2.5 Billion \autocite{noauthor_off_2021}.

NFTs were introduced by Ethereum \autocite{wood_ethereum_2014} as an improvement proposal \autocite{noauthor_eip-2309_nodate, noauthor_erc_nodate} in the \gls{erc}-721 standard \autocite{noauthor_erc-721_nodate}. This allows anyone to implement a Smart Contract with the ERC-721 standard and let people mint NFTs as well as, keep track of the tokens produced by it. This allows the created tokens to be validated.


\subsection{Smart Contracts \& ERC standards}

Smart Contracts are code that is running on the Blockchain. 3 of the notable ERC standards can be identified in table \ref{tab:erc-comparison}.

\begin{table}[htbp]
\caption{Comparison of ERC standards}
\begin{center}
\begin{tabular}{|p{0.19\linewidth}|p{0.19\linewidth}|p{0.19\linewidth}|p{0.19\linewidth}|}
\hline
\textbf{ERC-721} & \textbf{ERC-777} & \textbf{ERC-1155} & \textbf{ERC-20} \\ 
\hline
Non-fungible tokens & Non-fungible tokens \autocite{dafflon_eip-777_2017} & Semi-fungible tokens \autocite{prathap_semi-fungible_2021} & Fungible tokens \\ 
\hline
Each token is completely unique & A richer standard for fungible tokens, enabling new use cases and building on past learnings. Backwards compatible with ERC20. & Tokens begin trading as fungible tokens, then may end up being non-fungible in the long run & All coins of one kind are equivalent and hold the same value \\ 
\hline
CryptoKitties \autocite{cryptokitties_cryptokitties_nodate} &  & Concert tickets, gift vouchers, coupons & Crypto currencies - Bitcoin, ETH \\
\hline
\end{tabular}
\label{tab:erc-comparison}
\end{center}
\end{table}

\bigbreak
Each of the created tokens is unique from the other tokens created by the same Smart Contract, unlike fungible tokens which were introduced with cryptocurrencies and are denoted by the ERC-20 standard \autocite{noauthor_erc-20_nodate} on the Ethereum network. One Bitcoin can be swapped withn another Bitcoin, but each NFT will be unique.
Then, the deployed Smart Contract will be responsible to keep track of the tokens created by it on the network. A Smart Contract is a program that resides on the Ethereum network with a collection of code \& data \autocite{noauthor_introduction_nodate}.

For each NFT, the contact address \& unit256 tokenId are globally unique on any blockchain. This allows Decentralized Applications (DApps) \autocite{frankenfield_decentralized_nodate, noauthor_decentralized_2021} to take the tokenId and present the image/ asset that is identified by the particular NFT.

\subsection{\gls{nft} Marketplaces}
OpenSea, which was the first NFT marketplace is also considered to be the largest. In the attempt to become the "Amazon of NFTs", OepnSea raised \$23 million in a Series A \autocite{hackett_this_2021}, following a \$100 million raise in a Series B round, ended the company in a valuation of \$1.5 billion \autocite{dfinzer_announcing_2021, matney_nft_2021}. Open Sea saw nearly \$150 million in sales in the month of June.
These marketplaces are set to increase access to the digital goods industry \autocite{chevet_blockchain_2018}.

An NFT purchased on an Ethereum marketplace can be traded on any other Ethereum marketplace for a completely different NFT. Creators don't necessarily need to sell their NFT on a market. They can do the transaction peer-to-peer, completely secured by Blockchain. No one is needed to intermediate and an owner isn't locked onto any platform \autocite{noauthor_erc-721_nodate}.

\section{Motivation to explore how to Recommend \gls{nft}s}

\gls{recsys}s play a significant role in the resolution of the problem of information overload \autocite{cheng_hybrid_2020}. In order to provide ideal recommendations to a user, it is important to understand the user's thought process as well as other factors that affect a decision to trade.

Recommendation Systems have been driving engagement and consumption of content as well as items on almost every corner of the internet over the last decade.

These systems help users identify relevant items on an online platform. When users are recommended relevant items, it enables businesses in growing their revenue. 35\% of Amazon’s revenue \autocite{naumov_deep_2019} \& 60\% of watch time on YouTube \autocite{noauthor_recommendations_nodate} comes from recommendations. 75\% of Netlfix viewer activity \cite{vanderbilt_science_nodate} was also said to come from recommendations back in 2013.

Therefore, it is clear that the use of a recommendation system that is catered toward the needs of potential \gls{nft} owners will help increase sales of \gls{nft}s, driving forward the adoption of this technology

\bigbreak
Since generating relevant recommendations are highly important for many business use-cases and the \gls{nft} domain is seeing a booming acceptance with a bright future ahead, this work is expected to add value to the progression of advancements \& accessibility related to the domains of \gls{nft}s, Blockchain \& Recommendation Systems.

 In this research, the author attempts to identify features that could be considered for recommending NFTs and the importance of using multiple feature sets and algorithms to recommend relevant items.

\section{Value-driving factors of \gls{nft}s}

\subsection{Benefits of NFTs for creators, collectors \& buyers}
NFTs have a feature to allow a creator to make a certain percentage as royalty whenever the NFT is transferred to a new buyer. Since the items can be verified on the Blockchain, it also ensures that the original creator of the NFT can be tracked down and given due credit, on any date in the future, no matter how many wallets it gets passed through \autocite{chevet_blockchain_2018}. Apart from the fact that a buyer can claim the right of ownership of the original item, they also get to financially support the creator. Ultimately, NFTs may gain value over time due to their scarcity. This gives collectors an additional advantage of being able to sell it for a higher price later on.

Creators of NFTs can also create "shares" for their NFT. This allows investors and fans to own a portion of an NFT without having to purchase the entire thing \autocite{noauthor_erc-721_nodate}.

\subsection{Pricing of \gls{nft}s}

When considering the ownership desire of \gls{nft}s, it is understood that the increase in the price of an \gls{nft} has the possibility of being a factor to be considered when making a purchase.

The very first study done examining the pricing of NFTs suggests that \emph{"prospects for future studies are potentially limitless, as at the beginning of any new market"} \autocite{dowling_fertile_2021}. As a future study, the author has suggested identifying if there's a fundamental model that drives the price determination in NFTs.

\begin{quote} 
\centering 
\emph{"The value of an NFT is entirely determined by what someone else is willing to pay for it."}
\\
\raggedleft
\autocite{conti_what_2021}
\end{quote}

The value of an NFT has been identified to be heavily reliant on the public's acceptance of the item. Demand is expected to drive price rather than technical, or economic indicators which are the usual factors that affect stock prices and investor demand.

\begin{quote} 
\centering 
\emph{"Ultimately owning the real thing is as valuable as the market makes it. The more a piece of content is screen-grabbed, shared, and generally used the more value it gains. Owning the verifiable real thing will always have more value than not."}
\\
\raggedleft
\autocite{noauthor_erc-721_nodate}
\end{quote}

In addition to gaining value, due to the "non-fungible" nature of the item, it cannot be replicated. Similar to a Mona Lisa painting, popularity helps improve the value of the original, and only the original is identified as the truly original painting with immense value, even though anyone can Google and get a copy of the painting.

It is understood that \gls{nft}s have very little spill-over with other Crypto assets. However, knowing Crypto price prediction models is important since Wavelet coherence analysis indicates a co-movement between these two markets \autocite{dowling_is_2021}.
These models can be used separately on each NFT asset to anticipate the pricing related to time, sales \& bids.

\section{Existing Work}

\subsection{NFT Collections Recommendation System}

Conderation of the use of a basic \gls{ml} technique called \textbf{Multiple Regression} with data gathered from OpenSea in a blog article on \emph{OpenSea} \autocite{noauthor_what_2020}.

This takes into account previous purchase patterns and NFTs held in wallets to predict whether another wallet carrying a similar combination is likely to own an NFT from a certain category in the future. The categories considered here are mostly collections created by specific well-known creators. Cryptokitties and ENS domains are a couple of examples of collections that have been taken into consideration.

As a final recommendation, this system is capable of presenting NFT categories. Since users can't purchase an entire category, they will have to go back to the process of picking which NFT to purchase in the recommended collection.

This doesn't take into consideration of current global trends and it will not take into account the creators' recognition. An NFT minted by Beeple or a major league like NBA is bound to capture more attention of buyers compared to an NFT minted by a person who hasn't gained any reputation in this space. The major concern regarding this system is that the user must either enter his preferences manually or provide his wallet key, which holds all of his owned assets, to get a recommendation from the system. Although getting a users' public key can by no means cause any threat of losing the \gls{nft}s, it can lead to a lack of privacy, which is a tradition that the people into crypto-related assets have a tendency to be concerned about.

\subsection{Data Mining NFT Data from OpenSea}
One recent study done on data mining and visualizing has made use of the OpenSea Assets \& Events APIs using Python \& Pandas to collect, visualize \& analyse NFT data on Meebits Collection \autocite{larva_labs_meebits_nodate} \gls{nft} sales \autocite{adil_moujahid_data_2021}.

This work analyzes the outputs of the following data in the dataset.
\begin{enumerate}
\item Top 10 Meebits Creators, Buyers \& Sellers
\item The total number of Meebit Creators and Owners
\item Stats about Bundle/Single Sales
\item Types of Payment Currencies
\item Total Number of Sales per Day
\item Total Sales per Day in ETH \& USD
\item Average, Max \& Floor Meebit Price per Day in ETH
\end{enumerate}

While this work helps a lot with data mining, cleaning, preprocessing the data, and identifying the best possible users to target from a business perspective, it's doesn't explore how recommendations can be generated using the available data for specific items within the dataset.

\subsection{What may be the reasons for the lack of research related to recommending \gls{nft}s?}

\begin{quote} 
\centering 
\emph{"Crypto has a founding tradition of emphasizing freedom and privacy. Maybe because of this prevailing cultural trend, the NFT space does not have many recommender systems."} 
\\
\raggedleft
\autocite{noauthor_what_2020}
\end{quote}

As mentioned in the same blog post, this tradition is also been identified as a reason why we have not yet seen much development related to Recommendation Systems in this space. Another reason could be the very recent spark in interest this domain has seen in recent times.


\section{Proposed Approaches for Recommendations}

When conducting a requirement survey prior to building the prototype, the author understood that there was a clear necessity for \gls{nft} creators, buyers \& sellers to find items based on traits of an \gls{nft}. Traits are the properties that describe whatever that is contained in the image/ \gls{nft} asset.

Recommending items using \gls{nft} traits was attempted by the author in 2 different ways.

The \textbf{\textit{Bored Ape Yacht Club}}'s 10,000 \gls{nft}s were used to generate recommendations in this research. The Reference Id represented in the graphs is in the format of \textit{\gls{nft} Contract Address - Token Id}

\subsection{Trait Similarity Content-based Recommendations Approach}

The trait type and value were combined as lowercased strings to create a single string that would be unique even if \gls{nft}s from multiple collections were used to generate the cosine similarity matrix. A Count-Vectorizer was used to get a vectorized similarity score between all items considered for recommendations. The reason for choosing a Count-Vectorizer over a Tf-Idf Vectorizer was because all traits were considered equally important, to calculate an aggregate similarity score of all traits per item.

The top 10 items that had the cosine similarity score of the reference item's traits were taken as the recommendations here.

\begin{figure}[htbp]
\centerline{\includegraphics[width=\linewidth]{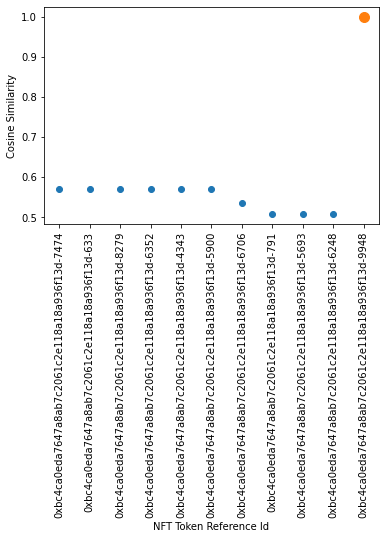}}
\caption{Trait Similarity Content based recommendations}
\label{fig:trait-content-output}
\end{figure}

\subsection{Trait Rarity based Recommendations Approach}
The rarer the traits are, the more valuable it would make an \gls{nft}. A rarity-score calculation method was introduced by rarity tools to calculate the total rarity-score of an item \autocite{raritytools_introducing_2021}.

\begin{figure}[htbp]
\begin{equation}
T_{r,t} = \sum^{Nt}_{t=1} \frac{1}{\left(\frac{c_{t}}{T_{N}}\right)}
\end{equation}
\caption{Equation to calculate the total trait rarity score of an \gls{nft} \autocite{raritytools_ranking_2021}}
\end{figure}

\noindent$T_{r,t}$ - Total rarity of a trait\\
$Nt$ - Total number of traits in the \gls{nft}\\
$c_{t}$ - Trait count of the chosen trait (number of occurrences in the collection)
$T_{N}$ - Total supply of \gls{nft}s in the collection

The absolute difference between the total rarities is calculated when an \gls{nft} from a collection is chosen. The lowest scoring items are recommended to the user. This gives the \gls{nft}s that may be as closely valuable as the initially chosen \gls{nft}.

The top 10 items that had the total rarity as close as possible to the reference item's rarity were taken as the recommendations here.

\begin{figure}[htbp]
\centerline{\includegraphics[width=\linewidth]{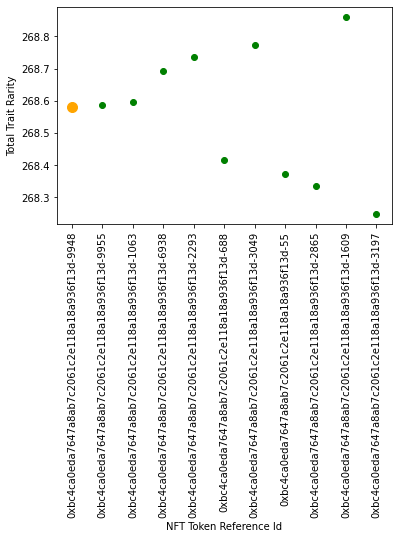}}
\caption{Total Rarity based recommendations}
\label{fig:total-rarity-output}
\end{figure}

\section{Evaluation}
The perfect measurement of evaluating a Recommendation System hasn't been the most straightforward. Since the Recommendation Systems introduced in this research attempt to recommend items in a very specific domain, the author decided to place the outputs produced by the two Recommendation Models in opposite graphs of measurements to visualize \& evaluate the produced outputs.

\begin{figure}[htbp]
\centerline{\includegraphics[width=\linewidth]{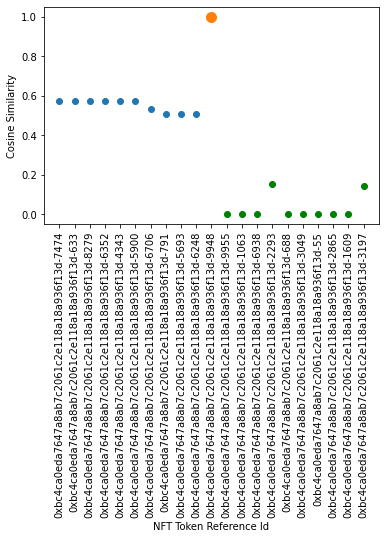}}
\caption{Cosine similarities of recommendations generated by both models}
\label{fig:combined-cosine}
\end{figure}

\begin{figure}[htbp]
\centerline{\includegraphics[width=\linewidth]{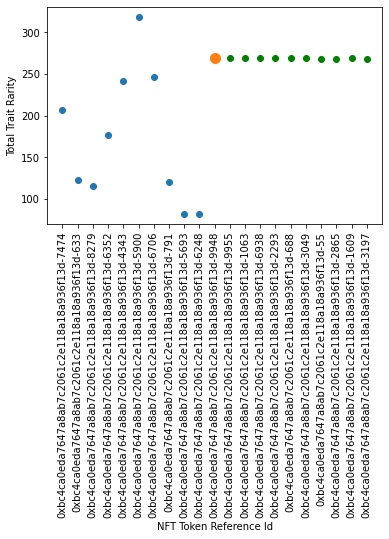}}
\caption{Total rarities of recommendations generated by both models}
\label{fig:combined-rarity}
\end{figure}

In the generated graphs represented in figures \ref{fig:combined-cosine} \& \ref{fig:combined-rarity}, the item used to generate recommendations for is represented in orange, the ones in blue were generated by the Trait similarity content based model, and the ones represented in green were generated by the Trait rarity model.
When taking a look at Figure \ref{fig:combined-cosine}, it's clear that none of the items recommended by the trait rarity model were even close to being content-wise similar.

When taking a look at Figure \ref{fig:combined-rarity}, the items recommended by the cosine similarity model were scattered around inconsistently.

As it has already been established that both these methods are necessary to be of value to a user when searching for items, it is clear that both the models are required to generate recommendations for a particular \gls{nft} using its traits.



\section{Conclusion}
In this research, the author set about exploring work related to utilizing available features of \gls{nft}s to recommend \gls{nft}s to users. After heading into the possibilities of recommending \gls{nft}s using their traits, the outputs produced by two suggested Recommendation Models were depicted in graphs. Finally, the importance of the two models were explained due to the outputs produced being contrastingly diverse. The foundation laid by the findings of this research could be built upon in future work to make the explorability and relevance of recommended items. This could help create better links between \gls{nft}s and users, resulting in better connections \& interactions among users as well as digital assets on the internet in the next decade.

\section{Future Enhancements}

If required to limit the recommendations produced, in order to recommend a fewer number of items with the most value to a user, it may be valuable to take the user's preference of each output into consideration. This could be done in a future study involving responses from human subjects.

Although the author initially planned to attempt recommending items using an \gls{lstm} \gls{dl} model since similar attempts have been researched with cryptocurrencies \autocite{ferdiansyah_lstm-method_2019}, it was understood that the very nature of uniqueness brought forwards by \gls{nft}s would not produce consistent results across collections or even items within the same collection. Even to attempt creating a bid price prediction model was challenging due to the lack of data and strict rate limits in the OpenSea \gls{api}. Closer to the completion of this research, the author came across an open dataset \autocite{zomglings_ethereum_2021} that may be usable for this purpose.

During the evaluation phase of the project, one of the feedback received regarding this was to attempt recommending \gls{nft}s using a dataset of price fluctuations in expensive, valuable physical artworks.

Furthermore, sentiment analysis is also proposed as future work to be combined with the \gls{lstm} method. This could be used to identify how public sentiment causes the value of crypto to adjust, in relation to past price fluctuations. This could be an interesting area to dive into since human-desire can be a major factor of consideration of the acceptance of a particular item.

Since the acceptance of \gls{nft}s seem to have a very common connection with social media such as \textit{Twitter}, \textit{Reddit} \& \textit{Discord communities}, it may be possible to recommend items using such data.

\section*{Acknowledgment}

The author of this paper acknowledges the guidance \& evaluation insights received for the project from Mr. Sharmilan Somasundaram \& Mr. Ragu Sivaraman.



\printbibliography 

\end{document}